\documentclass[sigconf]{acmart}

\AtBeginDocument{%
  }

\setcopyright{acmlicensed}
\copyrightyear{2025}
\acmYear{2025}
\acmDOI{XXXXXXX.XXXXXXX}

\acmConference[Conference acronym 'XX]{Make sure to enter the correct
  conference title from your rights confirmation email}{June 03--05,
  2018}{Woodstock, NY}

\acmISBN{978-1-4503-XXXX-X/18/06}

\graphicspath{{./images/}}
\usepackage{hyperref}
\usepackage{colortbl}
\copyrightyear{2025}
\acmYear{2025}
\setcopyright{rightsretained}
\acmConference[HT Adjunct 2025]{Adjunct Proceedings of the 36th ACM
Conference on Hypertext and Social Media}{September 15--18, 2025}{Chicago,
IL, USA}
\acmBooktitle{Adjunct Proceedings of the 36th ACM Conference on Hypertext
and Social Media (HT Adjunct 2025), September 15--18, 2025, Chicago, IL, USA}
\acmDOI{10.1145/3720533.3750059}
\acmISBN{979-8-4007-1533-4/2025/09}
\begin{document}

\title{Vocalize: Lead Acquisition and User Engagement through Gamified Voice Competitions}

\author{
Edvin Teskeredzic$^{1}$,
Muamer Paric$^{1}$,
Adna Sestic$^{1}$,
Petra Fribert$^{2}$,
Anamarija Lukac$^{2}$,
Hadzem Hadzic$^{1}$,
Kemal Altwlkany$^{1}$,
Emanuel Lacic$^{2}$
}

\affiliation{%
  \institution{$^{1}$Infobip BiH}
  \city{Sarajevo}
  \country{Bosnia and Herzegovina}
}

\affiliation{%
  \institution{$^{2}$Infobip Croatia}
  \city{Zagreb}
  \country{Croatia}
}

\email{{name.surname}@infobip.com}

\renewcommand{\shortauthors}{Teskeredzic, et al.}

\begin{abstract}
This paper explores the prospect of creating engaging user experiences and collecting leads through an interactive and gamified platform. We introduce Vocalize, an end-to-end system for increasing user engagement and lead acquisition through gamified voice competitions. Using audio processing techniques and LLMs, we create engaging and interactive experiences that have the potential to reach a wide audience, foster brand recognition, and increase customer loyalty. We describe the system from a technical standpoint and report results from launching Vocalize at 4 different live events. Our user study shows that Vocalize is capable of generating significant user engagement, which shows potential for gamified audio campaigns in marketing and similar verticals.
\end{abstract}

\ccsdesc[500]{Human-centered computing~User models}
\ccsdesc[300]{Human-centered computing~Natural language interfaces}
\ccsdesc[100]{Computing methodologies~Speech recognition}

\keywords{lead acquisition, user engagement, gamification, generative AI, speech processing}

\maketitle

\section{Introduction}

The active interaction of users with digital content has become a central pillar of success in modern business \cite{bag2022journey}. Active user engagement results in increased customer loyalty for existing clients~\cite{zheng2015building}, as well as the acquisition of new leads, often of high quality. However, generating user engagement through experiences that are both engaging and organic can be a challenging task for many organizations. One of the main hurdles in creating engagement, both for existing and new customers, is the lack of motivation and personalization when interacting with content~\cite{o2008user}. A possible solution to these problems comes in the form of user participation through gamified competitions.

\begin{figure}[tbp]
    \centering
    \includegraphics[width=\linewidth]{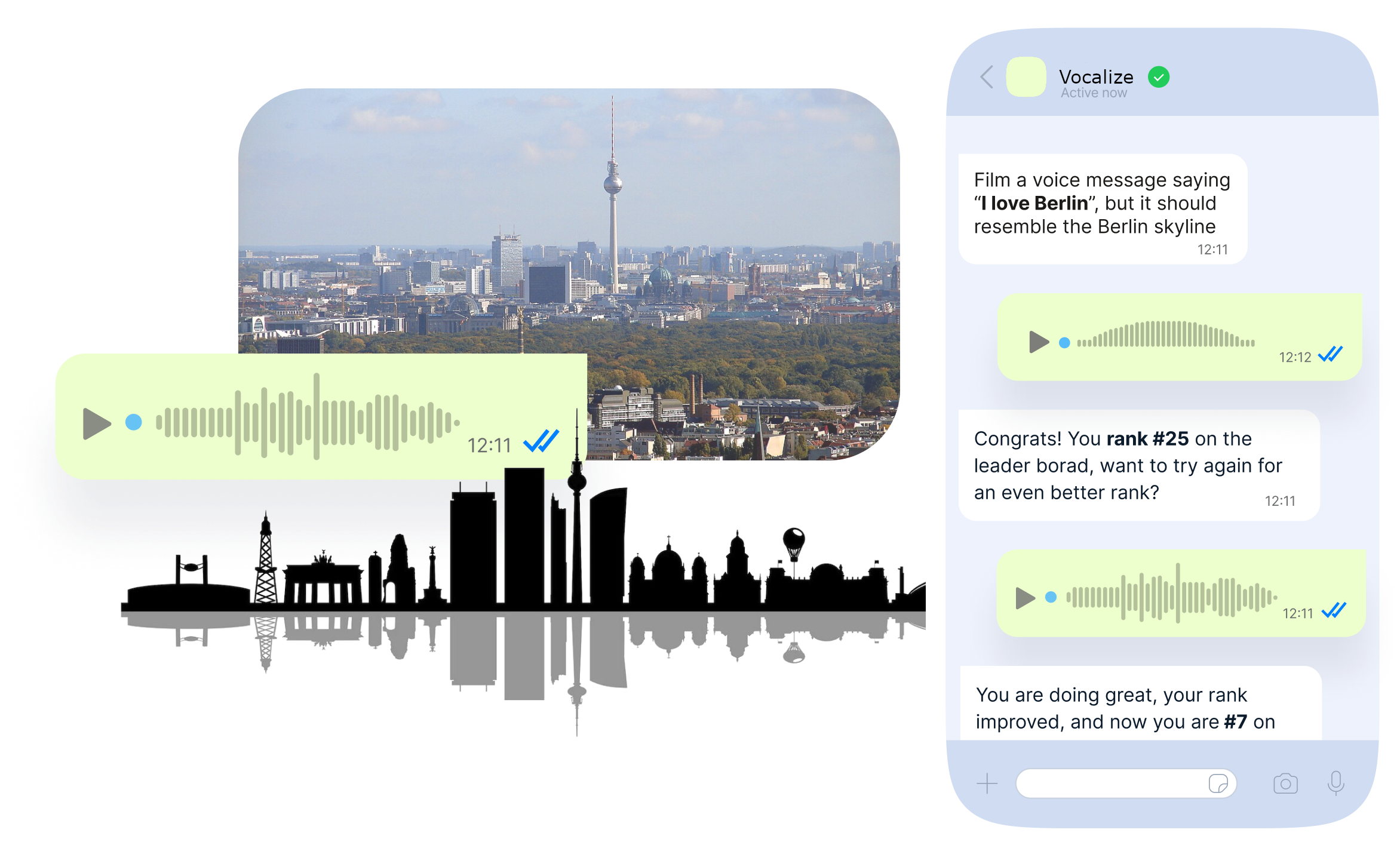}
    \caption{Example of a Vocalize campaign launched at WeAreDevelopers 2024.}
    \label{fig:vocalize}
    \Description{Screenshot of a Vocalize campaign interface at WeAreDevelopers 2024.}
\end{figure}

Voice-driven chatbot interfaces, when combined with gamification concepts, can transform how businesses acquire
leads and improve brand awareness. Such an approach makes interactions more engaging, and ensures personalized experiences that effectively capture attention and build lasting customer relationships~\cite{haleem2022artificial}. As demonstrated in prior research, gamified conversational agents have shown potential to boost motivation in educational contexts~\cite{khosrawi2023design}, while voice-based agents have proven effective in crafting immersive narratives~\cite{xu2022elinor}. Furthermore, frameworks leveraging dynamic language adaptation and conversational cues have been shown to foster emotional engagement and trust, maintaining user-centered interactions~\cite{joshi2024crafting,rayan2024exploring}.

\vspace{2mm}
\noindent
 \textbf{Contribution.} In this work we present \textit{Vocalize}\footnote{https://www.infobip.com/ai-hub/vocalize}, an end-to-end system for increasing user engagement and lead acquisition through gamified voice competitions. Vocalize offers a multimodal interface that allows users to communicate with its underlying chatbot using their voice and text. By combining audio-based gamification and generative AI-based content personalization, we show how to improve lead acquisition and increase customer engagement, which in turn can result in greater brand awareness~\cite{vo2022role}. We support our claims by analyzing the results obtained in lead acquisition and user engagement from 4 venues at which we launched Vocalize-supported campaigns.
\begin{figure*}[ht]
  \centering
  \includegraphics[width=0.8\textwidth]{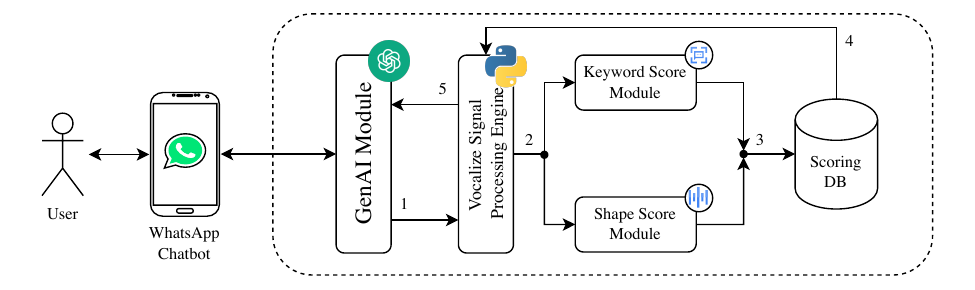}
  \caption{The underlying components of Vocalize. User audio inputs via WhatsApp are processed through keyword and shape scoring modules, with results enhanced by a generative AI and stored in a scoring database for feedback and leaderboard updates.}
  \label{fig:vocalize_schema}
\end{figure*}
\section{Vocalize}
For simplicity, we will refer to end users who interact with Vocalize as \textit{users}, while the term \textit{brand} is used for those designing and launching specific Vocalize campaigns. 
The main purpose of Vocalize is simple; enable brands to acquire leads of higher quality and have their users more engaged by providing them a gamified experience, with brands optionally handing out prizes to best ranked users.

Vocalize is built on top of WhatsApp\footnote{https://www.whatsapp.com/} as its underlying channel for communication. 
WhatsApp has been chosen as it is the most popular messaging application in terms of monthly active users - more than 3 billion~\cite{sinch_messaging_apps}.
To do this, we utilize Infobip's Answers platform\footnote{https://www.infobip.com/docs/answers} for integrating with WhatsApp as well as to access OpenAI's commercially available LLMs~\cite{openai2023gpt} to provide generative AI responses.

\vspace{2mm}
\noindent
\textbf{Competition outline.}
To initialize a competition, brands are only required to come up with a textual catch phrase and a desired image. 
For example, for a panoramic image of a city it could be the city skyline, or for an object it could be the edges of the object. 
Once users start interacting with the conversational agent over WhatsApp, they are provided with the predefined catch phrase and the outline of the image. Their task is then to record an audio message in which they repeat the catch phrase, but with an additional twist: the shape of the recorded audio message must also match the outline as close as possible. An example can be seen in Figure \ref{fig:vocalize} where the catch phrase "I love Berlin" was used in addition to the image that represents the skyline of Berlin. Here we further refer to the outline of that image as contour. Figure \ref{fig:vocalize} also shows how a user can improve their rank by recording an audio message that better matches the given contour. In summary, the main challenge of Vocalize in which users compete is to record an audio message that matches the campaign catch phrase in terms of content (spoken words), while also trying to match the shape of their audio recording (as visible in WhatsApp) with the target contour.

\vspace{2mm}
\noindent
\textbf{System architecture.} A complete overview of Vocalize is provided by Figure \ref{fig:vocalize_schema}. The WhatsApp chatbot interface enables users to communicate with Vocalize using natural language. If a user decides to participate in Vocalize's main challenge and sends an audio recording, it is forwarded to Vocalize's internal signal processing engine. The signal processing engine uses two separate components to formulate the score for the given user's audio recording: a keyword scoring module and a shape scoring module. The user's score is then logged to the database and the user is provided with feedback information. An example of such an interaction is shown in Figure \ref{fig:wa_chat}.

\begin{table*}[h!]
\caption{Statistics of the live events in 2024 where the user study was conducted. Larger events like WeAreDevelopers and Web Summit attracted significantly more voice recordings and longer interaction durations, while all events mostly demonstrated consistent message durations and engagement formats.}
\vspace{-2mm}
\tiny
\resizebox{\textwidth}{!}{
\begin{tabular}{l|l|l|l|c|r|c}
\multicolumn{1}{c|}{Event} & \multicolumn{1}{c|}{Date} & \multicolumn{1}{c|}{Target image} & \multicolumn{1}{c|}{Target phrase} & \multicolumn{1}{l|}{Voice recordings} & \multicolumn{1}{c|}{Total duration} & \multicolumn{1}{c}{Median duration} \\ \hline
WeAreDevelopers                                    & 17.07. - 19.07.                             & Skyline of Berlin                                       & I love Berlin                                               & $6,321$                                                               & $4$h $41$m &  $2.25$s \\
\rowcolor[HTML]{EFEFEF} 
KulenDayz                                          & 30.08. - 01.09.                             & Skyline of Osijek                                       & I love Kulen                                            & $1,257$                                                               & $46$m & $2.05$s \\
GOTO Chicago                                       & 21.10. - 22.10.                             & Skyline of Chicago                                          & Go to Infobip                                               & $1,216$                                                               & $1$h $42$m & $3.38$s   \\
\rowcolor[HTML]{EFEFEF} 
Web Summit                                         & 11.11. - 14.11.                             & Skyline of Lisbon                                      & I love Lisbon                                               & $3,662$                                                               & $2$h $22$m &  $2.69$s
\end{tabular}
}
\label{tab::events}
\end{table*}

\subsection{Chatbot Interface}

\begin{figure}[b!]
  \centering
  \includegraphics[width=0.75\columnwidth]{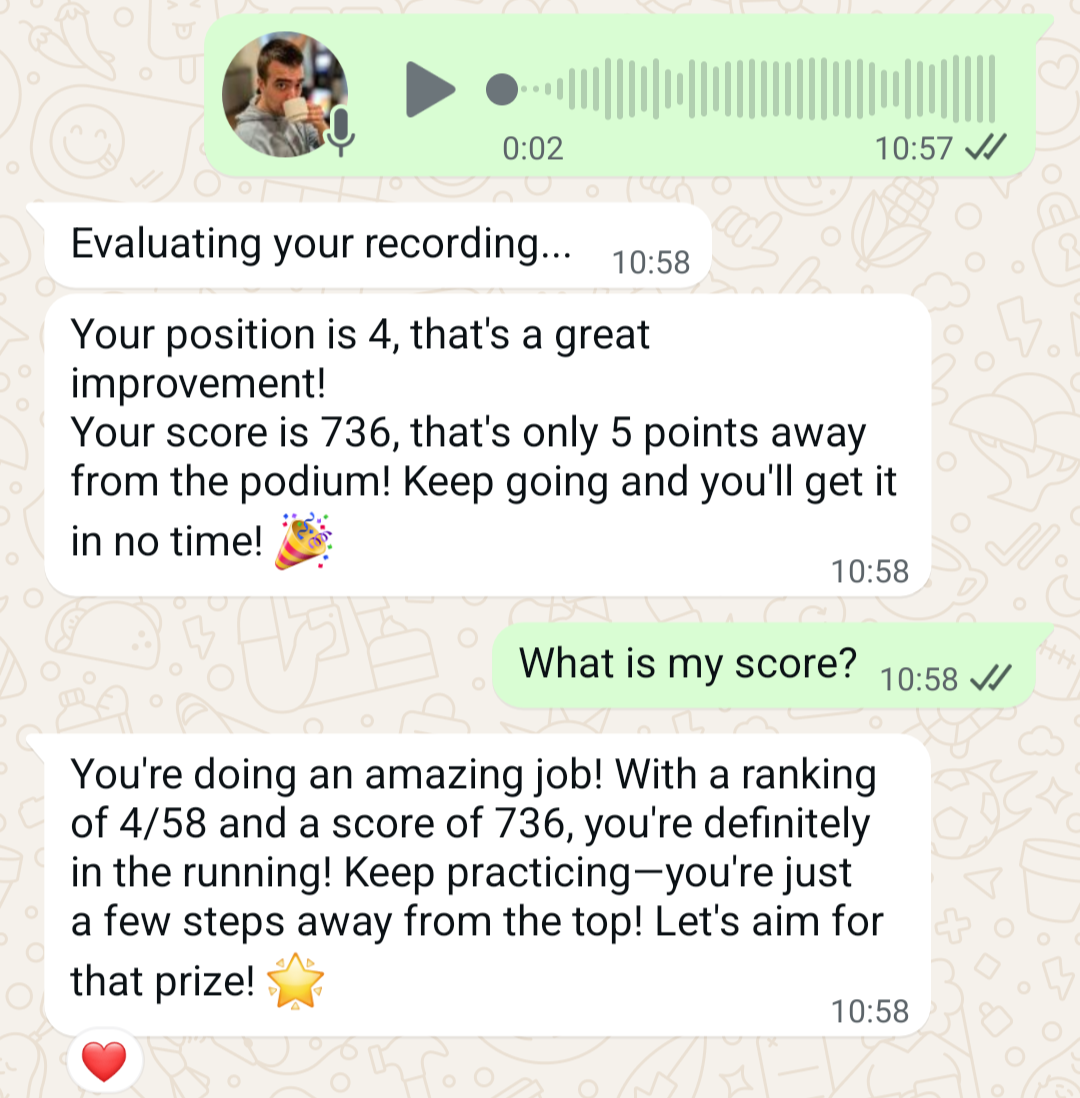}
  \caption{Users can interact with the system using natural language to receive scores, feedback, and personalized guidance in real time.}
  \label{fig:wa_chat}
\end{figure}

A typical way brands promote Vocalize is by advertising a QR code that initiates a conversation with Vocalize. This is useful for both virtual promotions, but also live venues as brands can print out the QR code on booths, posters or merchandise.

Once an interaction has been initiated, the conversational agent explains the rules of the game and collects contact information. As seen in Figure \ref{fig:wa_chat}, the generative AI model enables the user to interact with the system using natural language. For example, the user can ask the chatbot about their score or about the prizes without having to click through menus. If the user sends an audio message, it is automatically forwarded to Vocalize's underlying signal processing engine which assigns each recording a score. The score is returned, along with relevant metadata, e.g.: user's position on leaderboard, number of attempts made, next best score, etc. The generative AI module is used to polish the metadata and present it to the user in a more natural form that mimics a real conversation.

\subsection{Signal Processing Engine}
Vocalize relies on an underlying signal processing engine to assign scores to user recordings. It consists of two components, both of which produce an individual score for any given audio: keyword score and shape score. A Vocalize campaign can have one or both components active for scoring, which is defined by the brand.

\vspace{2mm}
\noindent
\textbf{Keyword scoring module.} This module rates how well the user's spoken content matches the reference catch phrase (i.e., a target phrase). The module outputs a score ranging from 0 (total mismatch) to 1 (perfect match). To determine what the user said, we obtain a transcript of the user's audio recording via automatic speech recognition. Depending on the campaign and brand requirements such as legal or privacy related considerations, we utilize an open-source solution, such as Whisper \cite{radford2023robust} or opt for a proprietary speech-to-text provider. The score itself is then computed by slightly modifying the Levenshtein algorithm for computing the distance between two strings \cite{levenshtein1966binary}:
\begin{equation}
    S = 1 - \frac{D}{L},
\end{equation}
where $D$ is the Levenshtein distance between the two strings and $L$ is the length of the longer of the two strings (target phrase and user transcript). If the two strings do not share characters in common, this score will equal 0, while if they are identical, the score will be 1, which directly matches our requirements.

\vspace{2mm}
\noindent
\textbf{Shape scoring module.} This module computes the visual similarity between the shape of the user audio recording (more precisely the envelope of its waveform) and the campaign contour.
Users only have visual access to the audio waveform as shown in WhatsApp's interface: segmented into 40 vertical bars. Thus, we adjusted the shape scoring module to accommodate for this. First, we segment both the contour and the audio waveform of the user temporally into 40 equally spaced segments. We compute the root mean square of each of these segments, thereby producing two temporal vectors of equal length (40) with non-negative values. Computing the similarity between these two vectors is then done using the dot product which, given the non-negativity of the vectors' elements, results in a value between 0 and 1, thereby matching our requirements and being in scale with the keyword scoring module.

Brands may choose which algorithm to use. We have not performed a detailed analysis of the performance differences between the two approaches and plan to do so in future work. Our speculation is that the first approach more closely resembles what humans consider to be "similar".

\subsection{Generative AI}
The use of the generative AI module as part of Vocalize is twofold:

\vspace{2mm}
\noindent
\textbf{Enhanced UX.} By leveraging intent detection, we enable the user to communicate with Vocalize using natural language. The user can query the system about updates regarding the competition (i.e. when does the competition close, what the user's current ranking is, etc.). We enable such communication by evaluating the embeddings of user messages, and classifying those into pre-determined intents. These intents are then mapped to Vocalize API calls, with the API response being fed to an LLM (in our case, OpenAI ChatGPT~\cite{openai2023gpt} family of models) in order to transform the JSON data into a natural language response, which gets presented to the user.

\vspace{2mm}
\noindent
\textbf{User engagement.} Vocalize uses a predefined conversational agent adapted through a system prompt~\footnote{The prompt message contained instructions to respond in an engaging, fun and gamified way and to provide the participant with instructions on how to compete and possibly win in the competition.}, when informing the user about their score after their attempt (i.e., after sending a voice message). The assistant is instructed to keep the user engaged, by offering encouragement, giving hints and tips, and using the current user's stats (i.e., number of attempts, best score, points to reach podium, etc.). Our goal is to increase user engagement while maintaining a fun, light-hearted tone when communicating with the participants.

\section{Demo and User Study}

\begin{table*}[h!]
\caption{Lead acquisition progression for all possible participants that sent an initial message (i.e., \textit{Potential leads}) to the chatbot. }
\vspace{-2mm}
\tiny
\resizebox{\linewidth}{!}{%
\begin{tabular}{l|c|c|c|c||c|c}
\multicolumn{1}{c|}{}                        & \multicolumn{1}{c|}{Potential leads} & \multicolumn{1}{c|}{Leads} & \multicolumn{1}{c|}{Participants} & \multicolumn{1}{c||}{Recurring participants} & \multicolumn{1}{c|}{Textual messages} & \multicolumn{1}{c}{Audio messages} \\ \hline
WeAreDevelopers                              & $430$              & 71.16\%              & 68.60\%              & 64.42\%              & 25\%              & 75\%                          \\
\rowcolor[HTML]{EFEFEF} 
KulenDayz                                    & $66$              & 71.21\%              & 69.69\%              & 69.69\%              & 21\%              & 79\%                          \\
GOTO Chicago                                 & $35$              & 71.43\%              & 65.71\%              & 60.00\%              & 13\%              & 87\%                          \\
\rowcolor[HTML]{EFEFEF} 
Web Summit                                   & $794$              & 59.57\%              & 42.32\%              & 27.58\%              & 65\%              & 35\%                         
\end{tabular}
}
\label{tab::dialog-progression}
\end{table*}

\begin{figure*}[h]
        \centering
        \includegraphics[width=0.9\textwidth]{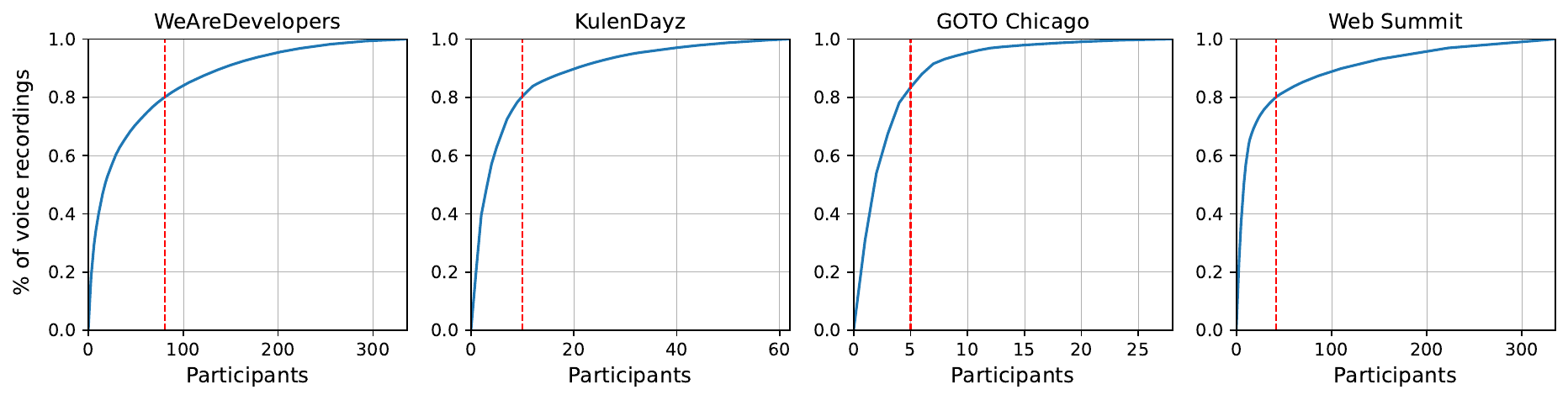}
        \vspace{-2mm}
    \caption{Cumulative frequency of voice recordings. Roughly 80\% of all voice messages (red) were sent by 24.11\% of participants in case of WeAreDevelopers, 16.13\% in case of KulenDayz, 17.86\% in case of GOTO Chicago and 12.54\% of participants at Web Summit.}
    \label{fig:hmt}
\end{figure*}

In order to validate the efficacy of Vocalize, we conducted a user study across two larger (i.e., WeAreDevelopers and  Web Summit Lisbon) and two smaller (i.e., Kulendayz and GOTO Chicago) live events in 2024. Here, our aim was to demonstrate the benefits of using gamified audio competitions in terms of user engagement, lead acquisition, and user experience.

 As seen in Table \ref{tab::events}, each competition incorporated unique target phrases aligned with the event theme as well as the skyline of the respective city of the event as the target image (also seen in Figure \ref{fig:vocalize}). The duration of the audio messages between all events ranged from $0.5$ to $16.8$ seconds. In addition, larger events did attract more participation with respect to the number of voice recordings and the total time of all voice messages recorded.
The overall performance of each voice-based competition is reported in Table \ref{tab::dialog-progression}. The reported proportion of audio messages reflect the number of voice recordings shown in Table \ref{tab::events}. Overall, we see a high participation rate once a user has sent an initial message (e.g., a \textit{potential lead} becomes a \textit{lead} once the user registration for the competition is finished, and a \textit{participant} if at least one voice recording has been sent). 
The progression from \textit{potential leads} to \textit{recurring participants} highlights the differences in user engagement across events. The smaller events demonstrated strong transition probabilities. In contrast, the larger events exhibited more varied outcomes. While WeAreDevelopers maintained high levels of recurring participation, Web Summit showed a notable drop-off. That is, at Web Summit we see a much higher proportion of textual messages when compared to voice recordings.
We hypothesize that the reason for such a behavior lies in the chatbot flow that was additionally extended for Web Summit - a Retrieval-Augmented Generation (RAG) component \cite{lewis2020retrieval} that invited participants of the conference (i.e., without the need of completing a registration) to learn more about Vocalize. Although, still fulfilling the purpose of brand awareness, such an addition may have inadvertently directed participants away from the competition and reduced the prominence of audio-based engagement.

\vspace{2mm}
\noindent
\textbf{Participation Dynamics.}
With respect to individual user engagement, as shown in Figure \ref{fig:hmt}, a small subset of highly active participants contributes the most to the total volume of recorded voice messages. This almost matches with the Pareto Principle~\cite{sanders1987pareto} as the distribution of recorded voice messages does have characteristics of a power-law distribution. For example, the most active participant at WeAreDevelopers recorded $460$ voice messages, at KulenDayz it was $259$, for GOTO Chicago $381$ and $298$ at Web Summit.
Our findings suggest that voice-based competitions in combination with conversational agents can resonate strongly with customers at live events.

\section{Conclusion and Future Work}
In this paper, we presented Vocalize - an end-to-end system for building voice-enabled competitions, based around the concept of matching a voice recording to a given target phrase and image. Furthermore, we gave a detailed breakdown of the main components of Vocalize from a technical standpoint. To assess the impact on user engagement and lead acquisition, we further evaluated our method across four live events where we demonstrated a strong participation rate and conversion from potential leads to recurring participants.

 We plan
 to enable the use of Vocalize with multiple communication channels, as well as customization of campaigns for use in marketing, event management, and exhibitions. 
 Here we plan to also investigate how the introduction of other novel audio scoring mechanisms in voice-based competitions impacts user engagement. 
 In addition, we intend to explore whether age, cultural, and industry background play an important role in engaging with gamified user experience by conducting further user studies at events aimed for different backgrounds.

\bibliographystyle{ACM-Reference-Format}
\bibliography{references.bib}

\end{document}